\documentclass[aps,prl,twocolumn,superscriptaddress,floatfix,letter,longbibliography]{revtex4-2}
\let\olddiv\div
\usepackage{epsfig,amsmath,amssymb,color,amsfonts,physics,microtype}
\usepackage[bookmarks=true,colorlinks,citecolor=red]{hyperref}
\usepackage{dsfont}
\usepackage{subcaption}
\usepackage{wrapfig}

\newcommand{\daga}{^\dag}

\begin{document}
\title{Multipartite Entanglement in the Measurement-Induced Phase Transition of the Quantum Ising Chain}
\author{Alessio Paviglianiti}
\author{Alessandro Silva}
\affiliation{International School for Advanced Studies (SISSA), via Bonomea 265, 34136 Trieste, Italy}

\begin{abstract}
    External monitoring of quantum many-body systems can give rise to a measurement-induced phase transition characterized by a change in behavior of the entanglement entropy from an area law to an unbounded growth. In this Letter, we show that this transition extends beyond bipartite correlations to multipartite entanglement. Using the quantum Fisher information, we investigate the entanglement dynamics of a continuously monitored quantum Ising chain. Multipartite entanglement exhibits the same phase boundaries observed for the entropy in the post-selected no-click trajectory. Instead, quantum jumps give rise to a more complex behavior that still features the transition, but adds the possibility of having a third phase with logarithmic entropy but bounded multipartiteness.
\end{abstract}

\maketitle

\it Introduction \rm --- In recent years, entanglement has become a key tool in condensed matter and statistical physics~\cite{amico2008}. Interesting quantum phenomena, such as quantum criticality or topological order, often give rise to peculiar behavior of entanglement too. For this reason, entanglement now plays an important role in understanding and classifying quantum many-body systems. This approach has lead to numerous theoretical insights, such as probing quantum phase transitions~\cite{calabrese2004,laflorencie2016}, understanding thermalization~\cite{kaufman2016}, and extracting information on topological properties~\cite{hamma2005,kitaev2006,levin2006}. In addition to its theoretical significance, entanglement is a fundamental resource of practical experimental use. For example, entangled states can enhance the precision of phase estimation in quantum metrology~\cite{pezze2014,pezze2018}, and are essential for implementing most quantum computing protocols, with relevant applications in quantum cryptography~\cite{gisin2002,horodecki2009}, optimization~\cite{santoro2006,albash2018}, and simulation~\cite{georgescu2014}.

Measurement-induced phase transitions~\cite{li2018,skinner2019,li2019,bao2020,ippoliti2021,coppola2022} are a notable example of entanglement as a diagnostic tool for quantum criticality in a dynamical setting. These recently discovered dynamical phase transitions occur when unitary many-body dynamics is punctuated by local measurements, and they feature a change in the scaling of the long-time entanglement entropy. Specifically, the rate of measurements performed on the system drives a transition from an area law phase to an entangling phase, typically with either volume law~\cite{choi2020,turkeshi2020,tang2020,boorman2022,legal2022,sierant2022bis} or logarithmic~\cite{alberton2021,turkeshi2021,szyniszewski2022,botzung2021} entanglement entropy depending on the model. The two phases have also been related to a qualitative change in the dynamical purification of a mixed state~\cite{gullans2020,gopalakrishnan2021}. Many features, including critical exponents~\cite{skinner2019,zabalo2020,turkeshi2022ter}, hints of conformal symmetry at criticality~\cite{jian2020,block2022,sharma2022}, and connections to the percolation universality class~\cite{skinner2019,lunt2021,sierant2022}, suggest that at least some versions of this phenomenon can be traced back to standard second order phase transitions. Despite all this progress, an exhaustive characterization of the transition is  still missing. For instance, it is still unclear whether there exists a local order parameter, and what would it look like in that case. In addition, even though some proposals have been put forward~\cite{goto2020,noel2022,buchhold2022,koh2022}, observing the transition in an experimental implementation remains an extremely challenging task due to the exponential complexity of post-selecting quantum trajectories.
\begin{figure}[ht]
\centering
 \includegraphics[width=\linewidth]{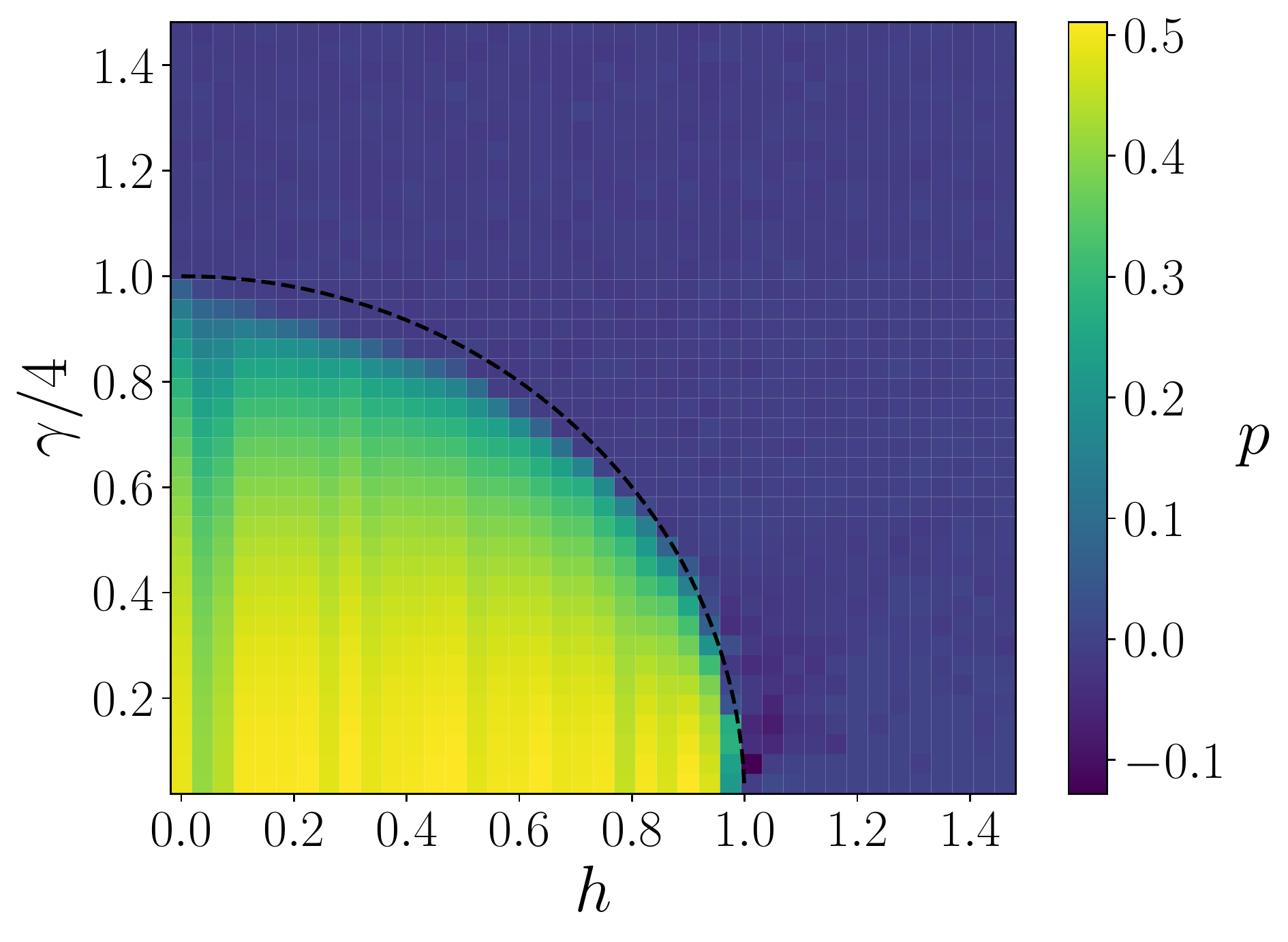}
 \caption{Exponent $p$ of $f_Q\sim L^p$ as a function of $h$ and $\gamma$ in the no-click limit. The dashed curve corresponds to the critical line $\gamma_c(h)$ that separates the gapped and gapless phases. The exponent is extrapolated by fitting data for $L=40 \olddiv 170$.}
 \label{f:phase_diagram}
\end{figure}

While the entanglement entropy effectively diagnoses measurement-induced transitions, it only probes bipartite entanglement, while entanglement itself is a more complex concept~\cite{amico2008,horodecki2009,hofmann2014,toth2012}. In particular, the structure of multipartite correlations is potentially richer and could give new insight for the characterization of the different phases. It is then natural to investigate whether monitored systems manifest a transition also in their multipartite entanglement structure. Genuine multipartite entanglement in quantum many-body systems can be witnessed using the quantum Fisher information (QFI), which may also give information about criticality and the associated order parameter~\cite{toth2012,hyllus2012,pezze2017,brenes2020}. This quantity has the advantage of being more accessible than the entanglement entropy in experimental realizations, as it can be obtained through measurements of dynamical susceptibilities~\cite{hauke2016,dealmeida2021}. Thus, even though the post-selection problem still remains, a characterization of the phase transition based on the QFI would represent a step towards its explicit experimental observation.

In this Letter, we investigate multipartite entanglement in a continuously monitored quantum Ising chain subject to  quantum jumps. We start by analyzing the so-called no-click limit, which has been found to qualitatively capture the main entanglement features of the measurement-induced transition~\cite{turkeshi2021,legal2022,zerba2023}, and we prove that the  QFI manifests the same phase diagram as the entropy. In the logarithmic phase, the QFI density features power law growth $f_Q \sim L^p$, corresponding to growing multipartiteness with system size, whereas it remains bounded in the area phase (see Fig.~\ref{f:phase_diagram}). We also consider the full dynamics with quantum jumps, revealing that in general the behavior of the QFI is more complex. While the 
density of QFI  still appears to grow as a power law in some parameter range within the logarithmic phase, its behavior at low values of the measurement rate is intensive. This and other observations show that the no-click limit does not capture all aspects of the model, and introduces the possibility of having a new phase with logarithmic entanglement entropy but bounded multipartite entanglement.

\it Model, measurement protocol, and QFI \rm --- Below, we consider measurement induced phase transitions in a quantum Ising chain in transverse field
\begin{equation}\label{hermitian_hamiltonian}
    \hat{H}_0 = -J\sum_j \hat{\sigma}_j^x\hat{\sigma}_{j+1}^x - h\sum_j\hat{\sigma}_j^z,
\end{equation}
with $L$ lattice sites and periodic boundary conditions. Throughout this Letter, we set $J=1$. Within the formalism of projective operator-valued measures~\cite{wiseman1996,ahnert2005}, we characterize entirely the measurement protocol by assigning suitable Kraus operators $\hat{A}_m$, $m=1,\dots, M$, satisfying $\sum_m \hat{A}\daga_m \hat{A}_m = \hat{\mathds{1}}$. In detail, given a state $\ket{\psi_t}$, the evolved state $\ket{\psi_{t+dt}}$ is obtained by applying a projector $\hat{A}_m$ to $\ket{\psi_t}$ and restoring the norm to $1$. The choice of the Kraus operator is performed randomly with probabilities set by $p_m = \bra{\psi_t}\hat{A}_m\daga \hat{A}_m\ket{\psi_t}$. 
In our case, we assume to measure the $z$-component of each spin randomly and independently of all others with a fixed rate $\gamma$. Since the full protocol can be broken down into single-site measurements, we use the local Kraus operators
\begin{subequations}
    \begin{equation}
    \hat{A}^{(0)}_j = (\hat{\mathds{1}}-\hat{L}_j) +\sqrt{1-\gamma dt}\,\hat{L}_j,
    \end{equation}    
    \begin{equation}
    \hat{A}^{(1)}_j = \sqrt{\gamma dt}\,\hat{L}_j,
    \end{equation}
    \end{subequations}
where $\hat{L}_j = \frac{1}{2}(\hat{\mathds{1}}+\hat{\sigma}_j^z) = \hat{L}_j\daga = \hat{L}_j^2$. Here the operators $\hat{A}^{(1)}_j$ describe sudden projections of the spins onto states with up $z$-component. Using this measurement protocol, the dynamics of the system is ruled by the stochastic Schrödinger equation~\cite{dalibard1992,daley2014}
\begin{equation}\label{sse}
    d\ket{\psi_t} = -i\hat{H}dt \ket{\psi_t} + \sum_j d\xi_{j,t}\left(\frac{\hat{L}_j}{\sqrt{\bra{\psi_t}\hat{L}_j\ket{\psi_t}}}-1\right)\ket{\psi_t},
\end{equation}
where
\begin{equation}\label{hamiltonian}
    \hat{H} = \hat{H}_0 - i\frac{\gamma}{4}\sum_j\left(\hat{\sigma}_j^z-\bra{\psi_t}\hat{\sigma}_j^z\ket{\psi_t}\right)
\end{equation}
is a non-Hermitian Hamiltonian describing an effective non-unitary evolution in absence of jumps, whereas the functions $d\xi_{j,t}=0,1$ are increments of independent Poisson processes satisfying $\overline{d\xi_{j,t}} = \gamma dt \bra{\psi_t}\hat{L}_j\ket{\psi_t}$. For details on the derivation of Eq.~\eqref{sse}, we refer the reader to Ref.~\cite{turkeshi2021}.

The scope of our study is to investigate the QFI in the stationary state of the dynamics generated by Eq.~\eqref{sse}. When evaluated on pure states, the QFI of an observable $\hat{O}$ takes a simple form proportional to its variance, namely,
\begin{equation}\label{qfi}
F_{Q}[\hat{O}] = 4 \left(\bra{\psi}\hat{O}^2\ket{\psi} - \bra{\psi}\hat{O}\ket{\psi}^2\right).
\end{equation}
As shown in Refs.~\cite{toth2012} and~\cite{hyllus2012}, this quantity can be used to witness multipartite entanglement when $\hat{O}=\hat{O}[{\{\mathbf{n}_j\}}]=\frac{1}{2}\sum_j\mathbf{n}_j\cdot \hat{\mathbf{\sigma}}_j$, where $\mathbf{n}_j$ are unit vectors. In this case, the QFI takes the form
\begin{equation}\label{qfi_specialized}
   F_{Q}[\hat{O}[{\{\mathbf{n}_j\}}]] = \sum_{\alpha,\beta=x,y,z}\sum_{i,j}n_i^\alpha C^{\alpha,\beta}_{i,j} n_j^\beta,
\end{equation}
where $C^{\alpha,\beta}_{i,j} = \bra{\psi}\hat{\sigma}^\alpha_i\hat{\sigma}^\beta_j\ket{\psi}-\bra{\psi}\hat{\sigma}^\alpha_i\ket{\psi}\bra{\psi}\hat{\sigma}^\beta_j\ket{\psi}$ are connected spin-spin correlators. If the density of QFI $f_Q = F_Q/L$ is larger than some divider $k$ of $L$, then the state $\ket{\psi}$ contains $(k+1)$-partite entanglement. The strictest lower bound to multipartite entanglement is obtained by finding the unit vectors $\{\mathbf{n}_j \}_{\rm opt}$ that maximize $F_Q$. The optimization problem is equivalent to the finding of the ground state of a classical Hamiltonian $H_{\rm cl} = -F_Q[\hat{O}[{\{\mathbf{n}_j\}}]]$ with vector spin variables $\mathbf{n}_j$, where the correlation functions play the role of 2-body couplings.

\it No-click limit \rm --- Let us start our analysis from the no-click limit, namely, the specific quantum trajectory in which all $d\xi_{j,t}$ are zero at all times. The time evolution is purely determined by the non-Hermitian Hamiltonian, and no quantum jump occurs. At long times, the dynamics converges to a stationary state, which coincides with the vacuum state of the non-Hermitian quasiparticles that diagonalize $\hat{H}$ up to a pair of quasiparticles that do not affect the entanglement entropy of the system~\cite{suppl_mat, zerba2023}. Even though this trajectory is exponentially unlikely, it can provide information on what can be expected in generic realizations of the full dynamics. For instance, Ref.~\cite{turkeshi2022} shows that the no-click limit of our model manifests the entanglement transition from area to logarithmic law. The logarithmic scaling of the entanglement entropy in the stationary state is linked to the absence of a gap in the decay rate of elementary excitations. For $|h|<1$ and $\gamma<\gamma_c(h) = 4\sqrt{1-h^2}$, the imaginary part of the quasiparticle spectrum is gapless, and the entanglement entropy follows a logarithmic law. In contrast, it is gapped outside this region, and the entropy obeys an area law. 

Interestingly, the two phases also feature a difference in their correlation functions. The spin-spin correlators $C^{\alpha,\beta}_{i,j}$ decay exponentially with the distance $|j-i|$ in the gapped phase, whereas they have power law envelope, modulated by sine-like oscillations, in the gapless phase~\cite{suppl_mat}. This difference impacts the QFI in the two phases: as mentioned, the maximization of the QFI is mapped into the search of the ground state energy of a classical Hamiltonian $H_{\rm cl}$ in which the correlation functions set the interactions. In the gapped phase, all correlators are exponential, and thus $H_{\rm cl}$ is a short-range Hamiltonian; as a consequence, we have $F_Q^\text{max}\sim L$, and $f_Q^\text{max}$ is intensive. The situation is different in the gapless phase, where the power law decay of correlations opens up the possibility of a long-range $H_{\rm cl}$. If the correlation functions decay slowly enough, one may expect a super-extensive scaling of the QFI with the system size, resulting in $f_Q^\text{max} \sim L^p$ with $p>0$. Recalling the connection between the QFI density and multipartite entanglement, this implies that the degree of multipartiteness of entanglement is bounded in the area phase, whereas it diverges as $\sim L^p$  in the logarithmic phase.

\begin{figure}[ht]
\centering
 \includegraphics[width=\linewidth]{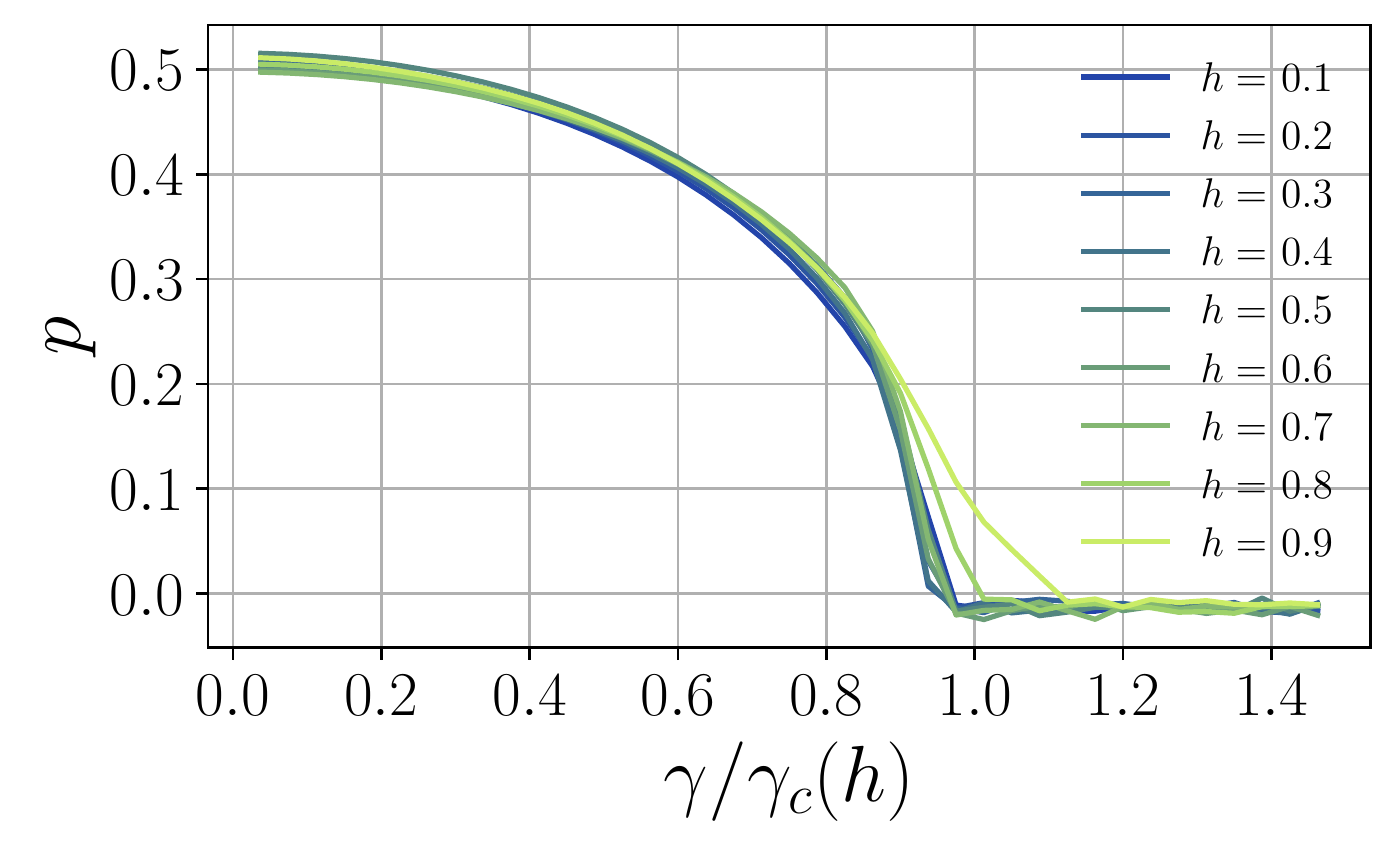}
 \caption{Exponent $p$ of $f_Q^\text{max}\sim L^p$ as a function of $\gamma/\gamma_c(h)$ for multiple values of $h$. The exponent decreases from $p\approx 0.5$ at $\gamma=0$ to zero at $\gamma\approx \gamma_c(h)$, and it appears to depend on $h$ only through $\gamma/\gamma_c(h)$.}
 \label{f:p_exponent}
\end{figure}

We now test numerically this hypothesis. We find the vacuum (steady) state by solving the model using the Jordan-Wigner map~\cite{mbeng2020}. The spin-spin correlators are computed using the methods described in Refs.~\cite{caianiello1952,barouch1971}, exploiting the Gaussian structure of the state~\cite{suppl_mat}. Finally, the maximization of the QFI is performed with a classical simulated annealing algorithm~\cite{bertsimas1993,ledesma2008}. For each choice of the parameters $h$ and $\gamma$, we evaluate the maximal QFI at different system sizes, and we fit the scaling of $f_Q^\text{max}$ to extrapolate the exponent $p$. Figure~\ref{f:phase_diagram} shows $p$ in the parameter space. Based on whether $p=0$ or $p>0$, we distinguish two phases, which overlap very well with the area and logarithmic phases diagnosed by the entanglement entropy. This result indicates that the entanglement transition in the no-click limit is witnessed by multipartite entanglement. Figure~\ref{f:p_exponent} shows the dependence of the exponent $p$ on $\gamma$ along vertical cuts of Fig.~\ref{f:phase_diagram}. These results suggest that $p$ might be a universal function of $\gamma/\gamma_c(h)$ for all values of $h$. We point out that the effective central charge of the entanglement entropy behaves similarly, being a function of $\gamma/\gamma_c(h)$ only~\cite{turkeshi2022}.

Surprisingly, while for translationally invariant systems 
the QFI is usually maximized by translationally invariant operators, this is not the case here. Specifically, in the gapless phase the optimal $\{\mathbf{n}_j\}_{\rm opt}$ are approximately aligned along the longitudinal $x$ direction, and  alternate between $+\mathbf{x}$ and $-\mathbf{x}$ with a wave vector $k = \pi - k^*$, where $k^*$ is the momentum at which the gap of the quasiparticle decay rate closes~\footnote{The definition of $k^*$ can actually be changed by adopting a different convention for the Jordan-Wigner mapping. See Supplemental Material for details.}. This is understood in terms of correlation functions. While we have no a-priori analytical prediction for its asymptotic behavior, we observe numerically that $C^{x,x}_{i,j}$ is the slowest-decaying spin-spin correlator, and it oscillates with a periodicity set precisely by $\pi - k^*$~\cite{suppl_mat}. This correlation function rules the leading order behavior of $f_Q^\text{max}$ with $L$, and thus the optimal configuration of unit vectors $\mathbf{n}_j$ must maximize its contribution in Eq.~\eqref{qfi_specialized}. Assuming the asymptotic ansatz $C^{x,x}_{i,j} \sim \cos((\pi-k^*)|i-j|)/|i-j|^\lambda$ with $\lambda<1$, which we find to be a good fit,
and considering for simplicity a periodic configuration $n_j^x = \cos((\pi-k^*)|i-j|)$, we obtain a contribution to the QFI that scales as $L^{2-\lambda}$: this yields a finite $p=1-\lambda > 0$.

Since the QFI is proportional to the variance of an observable, the operator $\hat{O}[{\{\mathbf{n}_j\}}_{\rm opt}]$ that maximizes it with a super-extensive variance can be seen as a local order parameter in what can be regarded as a critical region, both due to its entanglement entropy and its correlation functions. In this sense, $p$ may be interpreted as a critical exponent. For comparison, the order parameter of the quantum Ising chain $\hat{H}_0$ is the longitudinal magnetization $\sum_j\hat{\sigma}^x_j$, which is also the operator that maximizes the QFI  providing $f_Q^{\max}\sim L^{3/4}$ at the critical point~\cite{hauke2016}.
    \begin{figure}[ht]
\centering
 \includegraphics[width=\linewidth]{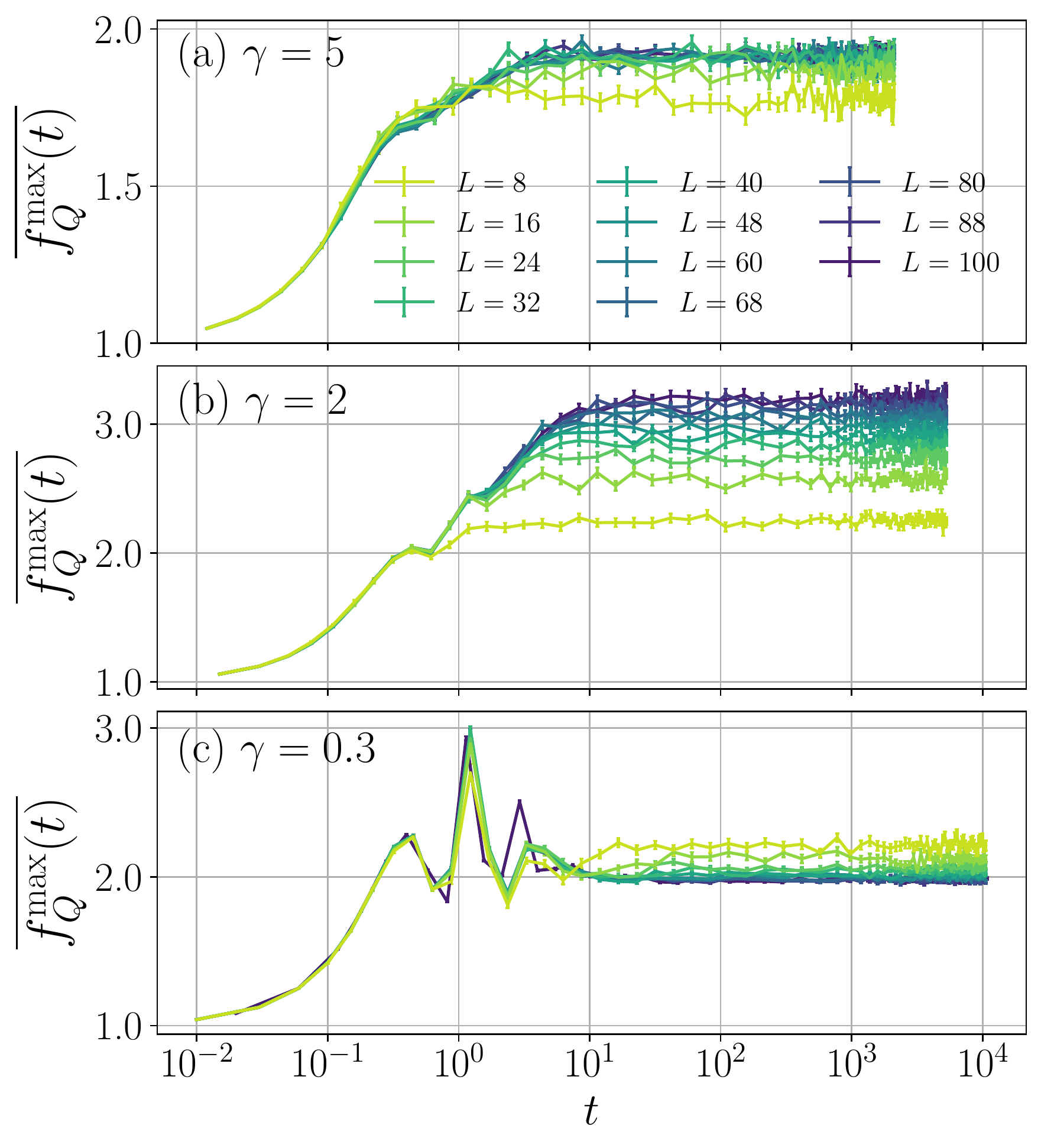}
 \caption{Dynamics of the disorder-averaged QFI density for $h=0.2$ and (a) $\gamma=5$, (b) $\gamma=2$, and (c) $\gamma=0.3$. The saturation value appears to be intensive at both large and small $\gamma$, whereas it grows with $L$ at intermediate $\gamma$.}
 \label{f:qfi_dynamics}
\end{figure}

\it Dynamics with quantum jumps \rm --- The results on the scaling of the QFI found in the no-click limit extend to the full dynamics produced by Eq.~\eqref{sse}, even though the phase diagram appears to be slightly modified. We start from a product state $\ket{\psi_0}$ with all spins along the positive $z$ direction, and characterize its dynamics using only the correlation matrices of Jordan-Wigner fermions~\cite{suppl_mat}, exploiting the preservation of the Gaussian nature of the state along each quantum trajectory. We compute the maximal QFI using simulated annealing, as in the no-click limit. Since this quantity depends on the previous history of quantum jumps, we repeat the procedure multiple times independently, and take a statistical average. To make a comparison, we also evaluate the entanglement entropy, defined as $S_\ell = -\Tr\left(\hat{\rho}_\ell\ln \hat{\rho}_\ell\right)$, where $\hat{\rho}_\ell$ is the reduced density matrix associated to a compact subsystem of $\ell$ spins.

We observe that already at the level of the average entanglement entropy the phase diagram does not appear to coincide with the one in the no-click limit presented in Ref.~\cite{turkeshi2022bis}; for instance, we observe a region of logarithmic entanglement entropy also for $h>1$ and sufficiently small $\gamma$~\cite{suppl_mat}. This could be either a finite size effect or, as suggested below by the QFI, an important difference between the no-click trajectory and the full dynamics~\cite{erratum,future_work}.
\begin{figure}
 \centering
 \includegraphics[width=\linewidth]
{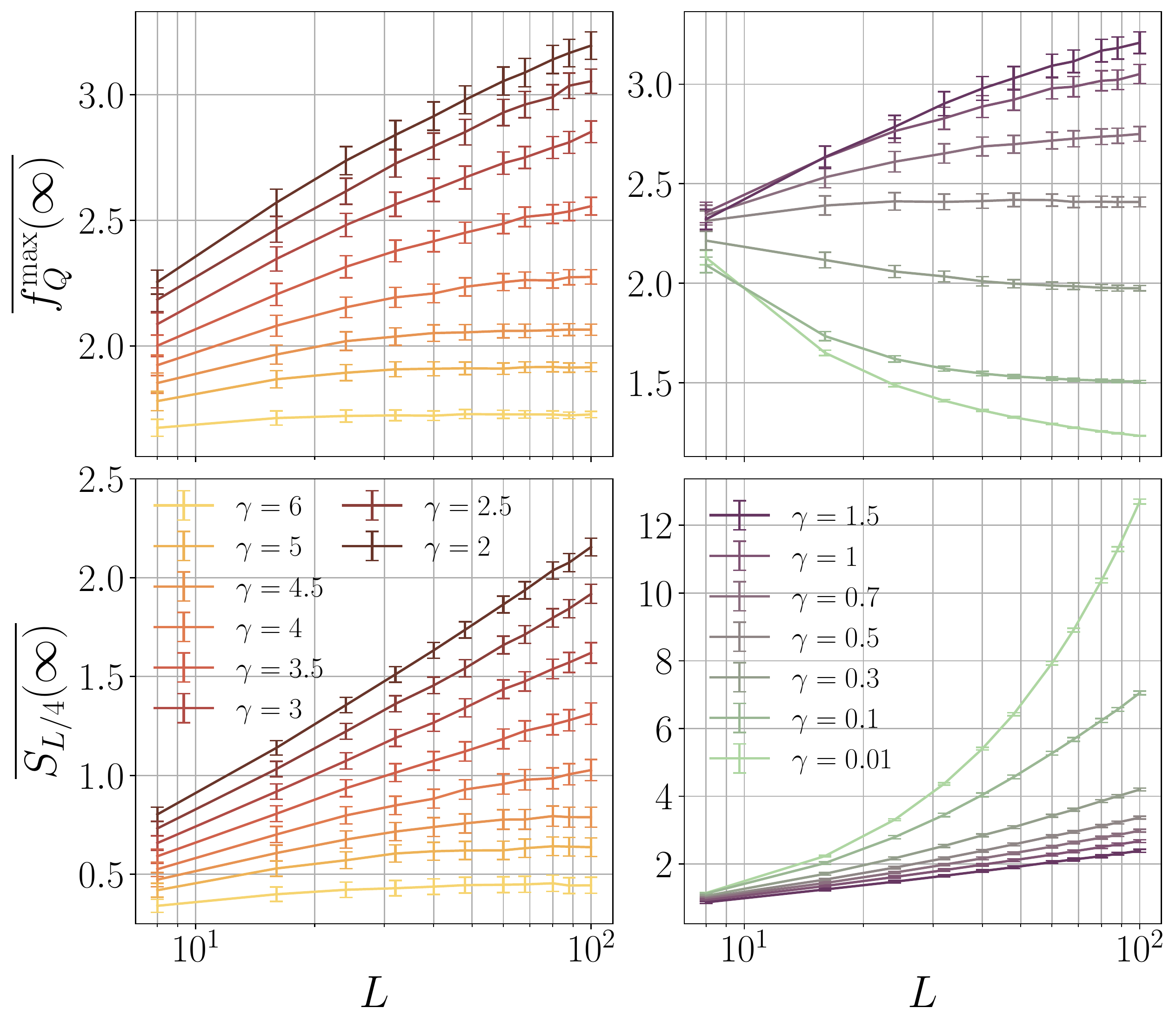}
 \caption{Disorder-averaged stationary QFI density and entanglement entropy, as functions of $L$, for $h=0.2$ (representative of other values~\cite{suppl_mat}) and multiple values of $\gamma$. Large (left panels) and small (right panels) values of $\gamma$ are presented separately to help visualization. The stationary values are evaluated as long-time averages.}
 \label{f:qfi_scaling}
\end{figure}

We now illustrate the results for the QFI for a given $h<1$, though similar results are obtained qualitatively for all $h$, even for $h>1$~\cite{suppl_mat}. At large values of $\gamma$, the average maximal QFI density saturates to an intensive value. In contrast, when $\gamma$ is reduced below $\gamma_c(h)$ ($\gamma_c \approx 4$ for $h=0.2$), $\overline{f_Q^\text{max}(\infty)}$ appears to grow indefinitely with the system size $L$. This is highlighted in Fig.~\ref{f:qfi_dynamics}(a)-(b), showing the dynamics of the average QFI density, and in the left panels of Fig.~\ref{f:qfi_scaling}, which portray the scaling of the stationary value with $L$. Our numerics suggest that the crossover of $f_Q$ from an intensive to a size-dependent value occurs at the same $\gamma_c$ at which $S_\ell$ transitions from area to logarithmic law, even though it is hard to determine the critical measurement rate precisely. The growth of the QFI density below $\gamma_c$ is consistent with a power law $\sim L^p$, as in the no-click limit.

When $\gamma$ is reduced further, we observe a new effect completely at odds with the no-click limit. The steady-state QFI density transitions back to an intensive value, as illustrated in Fig.~\ref{f:qfi_dynamics}(c) and in the right panels of Fig.~\ref{f:qfi_scaling}. The entanglement entropy shows also a contrasting behavior by developing a volume law  at small $\gamma$.  Our numerics suggest however that the entropy crossover takes place at lower $\gamma$ as compared to the QFI, and it could be interpreted as a finite-size effect occurring at $\gamma\sim 1/L$~\cite{alberton2021}, where the dynamics is approximately unitary and jumps are rare. Indeed, previous studies argue that the volume phase is unstable under any finite measurement rate $\gamma$ for free fermionic models~\cite{fidkowski2021,cao2019}.  These results may then indicate the presence of a third phase at low $\gamma$, featuring logarithmic entanglement entropy but only few-partite quantum correlations. 

We point out that often states with volume law entanglement entropy have a bounded QFI density, especially in the long-time unitary dynamics following a quantum quench~\cite{pappalardi2017}. This intuition might be consistent with the interpretation that the intensive scaling of $f_Q$ at small $\gamma$ is also a finite-size effect. While we cannot rule out this possibility, this seems to be at odds with our numerics showing that the average QFI density is smaller at larger values of $L$, where finite-size effects are less relevant. 

As in the no-click limit, the behavior of the QFI can be related to the shape of spin-spin correlation functions. However, since different trajectories have different optimal $\mathbf{n}_j$, identifying a local order parameter is harder if not impossible in this case. In addition, as seen below, a super-extensive average QFI does not imply that all trajectories have large multipartite entanglement. Focusing on single quantum trajectories in the long-time regime, where $\overline{f^\text{max}_Q(t)}$ has already reached saturation, we define the new correlators
\begin{equation}\label{c_tilde}
    \Tilde{C}^{\alpha,\beta}_\ell = \frac{1}{L}\sum_i\left|C^{\alpha,\beta}_{i,i+\ell}\right|,
\end{equation}
to study how the functions $C^{\alpha,\beta}_{i,j}$ decay with the distance $|j-i|$. The shape of all $\Tilde{C}^{\alpha,\beta}_\ell$ is exponential at large $\gamma>\gamma_c$, whereas they decay as power laws at smaller $\gamma$~\cite{suppl_mat}. This crossover already indicates a qualitative difference between the two regimes. The power law correlators also suggest that the entangling phase is an extended critical region, compatible with the observation of a logarithmic entanglement entropy. Given the analogy with the no-click limit, we expect that the average maximal QFI density diverges for $L\to \infty$ if these power laws decay slowly enough, namely, $\Tilde{C}^{\alpha,\beta}_\ell\sim|\ell|^{-\lambda_{\alpha,\beta}}$ with $\lambda_{\alpha,\beta}<1$. The exponents $\lambda_{\alpha,\beta}$ are however time-dependent, because individual quantum trajectories do not relax to stationary states. At any given (long) time, we estimate that a finite fraction of the trajectories in the ensemble of all possible random realizations has $\lambda_{\alpha,\beta}<1$, if $\gamma<\gamma_c$ but large enough~\cite{suppl_mat}, implying a divergent $\overline{f^\text{max}_Q(\infty)}$ as in Fig.~\ref{f:qfi_scaling}. In contrast, the fraction of quantum trajectories with $\lambda_{\alpha,\beta}<1$ goes to zero upon decrease of $\gamma$, explaining the return to a size-independent saturation value. We are unable to determine whether this fraction becomes exactly zero at low $\gamma$, which would indicate a strict phase transition, or remains small but finite. In any case, from the practical perspective of finite realizations samples and system sizes, both possibilities lead to an intensive 
$f_Q$.

\it Conclusions \rm --- In this Letter, we investigated the measurement-induced phase transition of a quantum Ising chain from the point of view of multipartite entanglement as witnessed by the QFI. In the post-selected trajectory without quantum jumps, the multipartiteness of quantum correlations changes from limited to extended, reproducing the same phase diagram obtained from the entanglement entropy. When quantum jumps are introduced, we still observe the transition, but a new region with bounded QFI density emerges at low $\gamma$. Our findings hint to the exciting possibility that a new phase might exist below a certain $\gamma'_c<\gamma_c$, featuring limited multipartite entanglement while maintaining logarithmic entanglement entropy. We further substantiate our analysis by investigating spin-spin correlation functions, and relating their behavior to the QFI.

We believe our study paves the way to future investigations on the role of multipartite entanglement in measurement-induced phase transitions. A topic of immediate interest is to establish whether or not the region with intensive QFI density at low $\gamma$ is indeed a stable phase. More broadly, we are still lacking a theoretical understanding of the whole phase diagram. The analysis of spin-spin correlators indicates that the logarithmic phase is a critical region, but an analytical prediction of this result is missing. Beyond our model, it will be interesting to study how multipartite entanglement behaves in different instances of the entanglement transition, including circuit models, systems with purely projective measurements, and, in particular, models with a volume phase.

\section*{Acknowledgements}

The authors are grateful to X. Turkeshi, M. Schiró, R. Fazio, and S. Pappalardi. A. P. also thanks D. Piccioni for providing useful advice on numerical methods.

\bibliography{bibliography}

\pagebreak
\widetext
\newpage
\begin{center}
\textbf{\large Supplemental Material}
\end{center}

\setcounter{equation}{0}
\setcounter{figure}{0}
\setcounter{section}{0}

\makeatletter
\renewcommand \thesection{S\@arabic\c@section}
\renewcommand{\theequation}{S\arabic{equation}}
\renewcommand{\thefigure}{S\arabic{figure}}

In order to complement the main text, in this Supplemental Material we provide additional discussion on our theoretical and numerical investigation. First, we present the diagonalization of the non-Hermitian Hamiltonian $\hat{H}$, and we define formally the stationary state of the no-click limit. We then illustrate in detail our numerical implementation of the dynamics with quantum jumps. Next, we discuss how to compute spin-spin correlation functions of Gaussian states. Moreover, we present numerical results on the correlators of our model, both in the no-click limit and in the long-time dynamics with jumps. Finally, we expand the results of the main text by showing the scaling of the steady-state QFI density and entanglement entropy for different values of the transverse field $h$, thus consolidating the generality of our findings.

\section{Diagonalization of the non-Hermitian Hamiltonian}
In the following, we provide a summarized description of the diagonalization of the non-Hermitian Hamiltonian, and we define the vacuum state. The discussion is very similar to the one presented in Ref.~\cite{turkeshi2022}, but we repeat it here, both to establish our notation, and because we adopt a different convention on the Jordan-Wigner map.

The effective Hamiltonian $\hat{H}$ remains integrable even if non-Hermitian, and can be diagonalized by mapping it to a free fermionic BCS model with a Jordan-Wigner transformation~\cite{mbeng2020}. In detail, we map spins into fermionic operators $\hat{c}_j$ and $\hat{c}\daga_j$ through the transformation
\begin{gather}\label{sigma_+}
    \hat{\sigma}^+_j = \frac{\hat{\sigma}^x_j+i\hat{\sigma}^y_j}{2} = e^{i\pi\sum_{i=1}^{j-1} \hat{n}_i} \hat{c}_j,\\ \label{sigma_-}
    \hat{\sigma}^-_j = \frac{\hat{\sigma}^x_j-i\hat{\sigma}^y_j}{2} = e^{i\pi\sum_{i=1}^{j-1} \hat{n}_i} \hat{c}\daga_j,
\end{gather}
where $\hat{n}_i = \hat{c}\daga_i\hat{c}_i$. For simplicity, we assume that $L$ is even. In terms of fermions, the effective Hamiltonian of Eq.~\eqref{hamiltonian} reads
\begin{equation}\label{hamiltonian_fermions}
    \hat{H} = - \sum_{j=1}^{L-1} \left(\hat{c}\daga_j\hat{c}_{j+1} + \hat{c}\daga_j\hat{c}\daga_{j+1} + \mathrm{h.c.}\right) + (-1)^{\hat{N}} \left(\hat{c}\daga_L\hat{c}_1 + \hat{c}\daga_L\hat{c}\daga_1 + \mathrm{h.c.}\right) + 2\left(h+i\frac{\gamma}{4}\right)\sum_{j=1}^L \hat{n}_j,
\end{equation}
where $\hat{N}=\sum_j\hat{n}_j$ is the total number of fermions. Here and in all following instances, any additive constant to the Hamiltonian is disregarded. While $\hat{N}$ itself is not conserved, its parity is a good quantum number. In our study, we work only with states in the even parity sector, and thus $\hat{H}$ is a BCS Hamiltonian with anti-periodic boundary conditions. We thus shift to momentum space by introducing the momenta $k$ satisfying $e^{i k L}=-1$, yielding $k = \pm \frac{2m-1}{L}\pi$, $m=1,\dots,L/2$. We define the Fourier-space fermionic operators
\begin{equation}
    \hat{d}_k = \frac{e^{-i\pi/4}}{\sqrt{L}}\sum_j e^{-i k j} \hat{c}_j,
\end{equation}
which allow us to rewrite the Hamiltonian as $\hat{H} = \sum_{k>0} \hat{H}_k$, where
\begin{equation}
    \hat{H}_k = \begin{pmatrix}
        \hat{d}_{-k} & \hat{d}\daga_k
    \end{pmatrix}\begin{pmatrix}
        2\cos k - 2h - i\frac{\gamma}{2} & -2\sin k\\
        -2 \sin k & -2\cos k +2 h + i\frac{\gamma}{2}
    \end{pmatrix}\begin{pmatrix}
        \hat{d}\daga_{-k} \\
        \hat{d}_k
    \end{pmatrix} = \begin{pmatrix}
        \hat{d}_{-k} & \hat{d}\daga_k
    \end{pmatrix}\begin{pmatrix}
        \epsilon_k & \Delta_k\\
        \Delta_k & -\epsilon_k
    \end{pmatrix}\begin{pmatrix}
        \hat{d}\daga_{-k} \\
        \hat{d}_k
    \end{pmatrix}.
\end{equation}
The two-particle Hamiltonian $\hat{H}_k$ acts on the manifold of states $\ket{0_k}$, $\ket{k}=\hat{d}\daga_k\ket{0_k}$, $\ket{-k}=\hat{d}\daga_{-k}\ket{0_k}$, and $\ket{k,-k}=\hat{d}\daga_k\hat{d}\daga_{-k}\ket{0_k}$, where $\ket{0_k}$ is the vacuum of fermions with momenta $\pm k$. It is immediately checked that $\hat{H}_k$ acts trivially on states with a single fermion, namely, $\hat{H}_k\ket{k} = \hat{H}_k\ket{-k} = 0$. As a consequence, these states have no dynamics, and thus we focus on states with even occupation of $\pm k$ fermionic modes.

The eigenvalues of the Hamiltonian $\hat{H}_k$ are found in opposite pairs $\pm \Lambda_k$, where
\begin{equation}\label{lambda_k}
    \Lambda_k = 2\sqrt{1-2h\cos k + h^2 -\frac{\gamma^2}{16} + i \frac{\gamma}{2}(h-\cos k)}.
\end{equation}
The complex square root requires the choice of a branch. For each $k$, we are free to define it in such a way that $\Lambda_k = E_k + i \Gamma_k$ (with $E_k, \Gamma_k\in \mathbb{R}$) has non-positive imaginary part $\Gamma_k\leq 0$. This choice is always possible, as eigenvalues come in pairs of opposite sign. The Hamiltonian takes the diagonal form
\begin{equation}
    \hat{H}_k = \Lambda_k\left(\hat{\Bar{\gamma}}_{-k}\hat{\gamma}_{-k} + \hat{\Bar{\gamma}}_k\hat{\gamma}_k\right)
\end{equation}
in terms of new fermionic operators
\begin{gather}
    \hat{\Bar{\gamma}}_k = \frac{-(\Lambda_k-\epsilon_k)\hat{d}\daga_{-k} + \Delta_k\hat{d}_k}{\sqrt{2\Lambda_k(\Lambda_k-\epsilon_k)}},\\
    \hat{\gamma}_k = \frac{-(\Lambda_k-\epsilon_k)\hat{d}_{-k} + \Delta_k\hat{d}\daga_k}{\sqrt{2\Lambda_k(\Lambda_k-\epsilon_k)}}.
\end{gather}
The choice of the branch of the square roots is irrelevant, as long as the same convention is picked for both operators. The diagonal fermions satisfy the canonical anticommutation relations $\acomm{\hat{\Bar{\gamma}}_k}{\hat{\gamma}_{k'}} = \delta_{k,k'}$, $\acomm{\hat{\Bar{\gamma}}_k}{\hat{\Bar{\gamma}}_{k'}} =\acomm{\hat{\gamma}_k}{\hat{\gamma}_{k'}} = 0$, but, differently from the Hermitian case, we have $\hat{\Bar{\gamma}}_k\neq \hat{\gamma}\daga_k$. Nevertheless, these can still be interpreted as creation and annihilation operators for non-Hermitian quasiparticles carrying complex energies $\Lambda_k$, and the operators $\hat{\Bar{\gamma}}_k\hat{\gamma}_k$ and $\hat{\Bar{\gamma}}_{-k}\hat{\gamma}_{-k}$ have the meaning of (non-conserved) number operators.

As mentioned in the main text, below the critical line $\gamma_c(h) = 4\sqrt{1-h^2}$ and for $|h|<1$ the imaginary part $\Gamma_k$ is gapless, i.e., it vanishes at $k^* = \arccos h$. From Eq.~\eqref{lambda_k}, we identify $k^* = \arccos h$. We point out, however, that this specific value depends on the convention adopted for the Jordan-Wigner mapping. In fact, an alternative to Eqs.~\eqref{sigma_+} and~\eqref{sigma_-} is to swap the definitions $\hat{\sigma}^+_j\leftrightarrow\hat{\sigma}^-_j$. The resulting Hamiltonian is the same as Eq.~\eqref{hamiltonian_fermions}, but with $h \to -h$. As a consequence, this alternative mapping yields $k^* = \arccos (-h) = \pi - \arccos h$.

We can finally introduce the vacuum state of non-Hermitian quasiparticles. For each $k$, we define $\ket{vac_{\, k}}$ as the state that is annihilated by $\hat{\gamma}_{\pm k}$, obtaining
\begin{equation}
    \ket{vac_{\, k}} = \frac{(\Lambda_k-\epsilon_k)\ket{0_k} - \Delta_k \ket{k,-k}}{\sqrt{|\Lambda_k-\epsilon_k|^2+\Delta_k^2}}.
\end{equation}
When acting on this state, the operators $\hat{\Bar{\gamma}}_{\pm k}$ add quasiparticles with complex energy $\Lambda_k$.  In particular, within the same parity sector, we may define the state with two non-Hermitian quasiparticles $\hat{\Bar{\gamma}}_k\hat{\Bar{\gamma}}_{- k}\ket{vac_{\, k}}$, which, together with $\ket{vac_{\, k}}$, spans the same space as $\ket{0_k}$ and $\ket{k,-k}$. Any initial state within this space can then be represented as
\begin{equation}
    \ket{\psi_k(0)} = \alpha \ket{vac_{\, k}} + \beta\,\hat{\Bar{\gamma}}_k\hat{\Bar{\gamma}}_{- k}\ket{vac_{\, k}},
\end{equation}
where $|\alpha|^2 + |\beta|^2 = 1$. Since we defined $\Im \Lambda_k$ to be negative, it follows from
\begin{equation}
    \ket{\psi_k(t)} = \frac{e^{-i\hat{H}_k t}\ket{\psi_k(0)}}{\sqrt{\bra{\psi_k(0)}e^{i\hat{H}\daga_k t}e^{-i\hat{H}_k t}\ket{\psi_k(0)}}}
\end{equation}
that $\ket{vac_{\, k}}$ is the steady state of the dynamics reached for $t\to \infty$, as the exponential factor $e^{2 \Gamma_k t}$ suppresses the relative weight of the state with quasiparticles. Considering all modes $k$, the overall vacuum state of the system is simply $\ket{vac} = \otimes_{k>0}\ket{vac_{\, k}}$.

\section{Numerical implementation of the dynamics}
We consider the evolution of an initial product state generated by the stochastic Schrödinger equation of Eq.~\eqref{sse}. Since the state remains Gaussian throughout the dynamics for any disorder realization, it is sufficient to evaluate its two-fermion correlation matrices as functions of time rather than track the full state $\ket{\psi_t}$. Specifically, all relevant information is stored within the matrices $C_{m,n}(t) = \bra{\psi_t}\hat{c}_m\hat{c}\daga_n\ket{\psi_t}$ and $F_{m,n}(t) = \bra{\psi_t}\hat{c}_m\hat{c}_n\ket{\psi_t}$. We derive evolution equations for $C(t)$ and $F(t)$ under both the non-unitary dynamics and the sudden quantum jumps, similarly to what is presented in the Supplemental Material of Ref.~\cite{turkeshi2022bis}.

In absence on jumps, the state $\ket{\psi_t}$ evolves according to a deterministic Schrödinger equation with Hamiltonian $\hat{H}$. In terms of the matrices $C(t)$ and $F(t)$, the dynamics can still be formulated in terms of first-order differential equations, which are derived in the following way. We explicitly compute the time derivatives
\begin{gather}\label{C_derivative}
    \partial_t C_{m,n}(t) = i\bra{\psi_t}\left(\hat{H}\daga \hat{c}_m\hat{c}\daga_n - \hat{c}_m\hat{c}\daga_n\hat{H}\right)\ket{\psi_t},\\\label{F_derivative}
    \partial_t F_{m,n}(t) = i\bra{\psi_t}\left(\hat{H}\daga \hat{c}_m\hat{c}_n - \hat{c}_m\hat{c}_n\hat{H}\right)\ket{\psi_t},
\end{gather}
where $\hat{H}$ is the Hamiltonian of Eq.~\eqref{hamiltonian}. It is fundamental to include also the constant term $i\frac{\gamma}{4}\sum_j\bra{\psi_t}\hat{\sigma}^z_j\ket{\psi_t}$, as it enforces the conservation of the norm $\braket{\psi_t}{\psi_t}=1$ at all times. The right-hand sides of Eqs.~\eqref{C_derivative} and~\eqref{F_derivative} can be worked out explicitly using Wick's theorem, and they finally yield
\begin{gather}
    \partial_t C(t) = -2i\left(\comm{\mathbb{H}_1}{C(t)} + \mathbb{H}_2 F\daga(t) + F(t) \mathbb{H}_2\right) + \gamma \left(C(t)^2 - F(t)F\daga(t) -C(t)\right),\\
    \partial_t F(t) = -2i\left[\acomm{\mathbb{H}_1}{F(t)} + \mathbb{H}_2 (\mathds{1}-C^T(t)) -C(t) \mathbb{H}_2\right] + \gamma \left[C(t)F(t) - F(t)(\mathds{1}-C^T(t))\right],
\end{gather}
where $\mathbb{H}_{1,2}$ are $L \times L$ matrices with non-zero elements
\begin{gather}
    (\mathbb{H}_1)_{m,m+1} = (\mathbb{H}_1)_{m+1,m} = -\frac{1}{2},\\
    (\mathbb{H}_1)_{L,1} = (\mathbb{H}_1)_{1,L} = \frac{1}{2},\\
    (\mathbb{H}_1)_{m,m} = h,
\end{gather}
and
\begin{gather}
    (\mathbb{H}_2)_{m,m+1} = -(\mathbb{H}_2)_{m+1,m} = -\frac{1}{2},\\
    (\mathbb{H}_2)_{L,1} = -(\mathbb{H}_2)_{1,L} = \frac{1}{2}.
\end{gather}
These matrices encode the contribution of the Hermitian Hamiltonian $\hat{H}_0$, which can be written (apart from a constant) as
\begin{equation}
    \hat{H}_0 = \begin{pmatrix}
        \hat{c}\daga_1 & \dots & \hat{c}\daga_L & \hat{c}_1 & \dots & \hat{c}_L
    \end{pmatrix}\begin{pmatrix}
        \mathbb{H}_1 & \mathbb{H}_2\\
        -\mathbb{H}_2 & -\mathbb{H}_1
    \end{pmatrix} \begin{pmatrix}
        \hat{c}_1 \\ \dots \\ \hat{c}_L \\ \hat{c}\daga_1 \\ \dots \\ \hat{c}\daga_L
    \end{pmatrix}.
\end{equation}

The discontinuous evolution produced by quantum jumps is implemented as follows. Suppose the jump occurs on site $j$. The starting state $\ket{\psi}$ is projected onto
$\ket{\psi'} = \frac{\hat{L}_j\ket{\psi}}{\sqrt{\bra{\psi}\hat{L}_j\ket{\psi}}}$, where $\hat{L}_j = \hat{\mathds{1}}-\hat{n}_j$ given our definition of Eqs.~\eqref{sigma_+} and~\eqref{sigma_-}. It follows that the correlation functions $C$ and $F$ are updated to
\begin{gather}
    C'_{m,n} = \frac{\bra{\psi}\hat{L}\daga_j\hat{c}_m\hat{c}\daga_n\hat{L}_j\ket{\psi}}{\bra{\psi}\hat{L}\daga_j\hat{L}_j\ket{\psi}},\\
    F'_{m,n} = \frac{\bra{\psi}\hat{L}\daga_j\hat{c}_m\hat{c}_n\hat{L}_j\ket{\psi}}{\bra{\psi}\hat{L}\daga_j\hat{L}_j\ket{\psi}},
\end{gather}
which can be evaluated using Wick's theorem. The result reads
\begin{gather}\label{C_jump}
    C'_{m,n} = C_{m,n} - \frac{C_{m,j}C_{j,n}}{C_{j,j}} + \frac{F_{m,j}F\daga_{j,n}}{C_{j,j}} + \delta_{j,m}\delta_{j,n},\\\label{F_jump}
    F'_{m,n} = F_{m,n} - \frac{C_{m,j}F_{j,n}}{C_{j,j}} + \frac{F_{j,m}C_{n,j}}{C_{j,j}}.
\end{gather}
It is worth noting that the $j$th rows and columns of both $C$ and $F$ are updated to $C'_{m,j} = C'_{j,m} = \delta_{m,j}$ and $F'_{m,j} = F'_{j,m} = 0$, independent of their previous values. In our implementation, we impose such conditions explicitly instead of using Eqs.~\eqref{C_jump} and~\eqref{F_jump} for these specific rows and columns, because we observed that this improves numerical stability.

In our numerics, we discretize time in steps of duration $\delta t$. At each step, we implement the non-unitary evolution by integrating the coupled differential equations for $C(t)$ and $F(t)$ using a Runge-Kutta algorithm of $5^{th}$ order. We then check whether a quantum jump randomly occurs or not, following the method adopted in Ref.~\cite{turkeshi2021}. The expected number of jumps on each site $j$ is equal to its jump probability $p_j = \gamma \delta t \bra{\psi_t}\hat{L}_j\ket{\psi_t}$, and thus the total expected number of jumps is $P = \sum_j p_j$. For sufficiently small $\delta t$, $P<1$ can be interpreted as the probability to have a jump on one of the lattice sites. We thus extract a random number $0\leq r\leq 1$ and we compare it to $P$. If $r>P$, no jump occurs. In contrast, for $r\leq P$ we implement a jump on a site $m$, which is chosen randomly among all sites using $p_j$ as relative probabilities. We point out that an alternative (but equivalent) implementation consists of checking if a jump occurs on each single site independently, which allows for multiple jumps within the same time step. If $\delta t$ is small enough, the two methods coincide, because processes involving $n$ jumps happen with probability proportional to $(\delta t)^n$. Regardless, both implementations produce the same average number of jumps $P$, which is the physically relevant quantifier of how frequently entanglement is reduced by a projective measurement. We checked numerically that these two approaches produce no notable difference in the quantities we investigated.

\section{Correlation functions}
We now show how to evaluate the connected spin-spin correlation functions $C^{\alpha,\beta}_{m,n}$. As a preliminary step, we briefly summarize how to evaluate the two-point fermionic correlation functions of a translationally-invariant Gaussian state in the generic form
\begin{equation}
    \ket{\psi} = \otimes_{k>0}\ket{\psi_k} = \otimes_{k>0}\left(u_k \ket{0_k} + v_k \ket{k,-k}\right),
\end{equation}
where $|u_k|^2+|v_k|^2 = 1$. In particular, the following discussion applies to the vacuum state of non-Hermitian quasiparticles. Owing to Wick's theorem, correlation functions can be expressed in terms of two-body correlators of fermionic operators. For convenience, we use the the Majorana operators $\hat{A}_j = \hat{c}\daga_j+\hat{c}_j$ and $\hat{B}_j = \hat{c}\daga_j-\hat{c}_j$ instead of the original $\hat{c}_j$ and $\hat{c}\daga_j$. Their correlations functions can be computed analytically (see Supplemental Material of Ref.~\cite{turkeshi2021} for the details), and the result is
\begin{gather}
    (M_{AA})_{m,n}= \bra{\psi}\hat{A}_m\hat{A}_n\ket{\psi} = \delta_{m,n} + \frac{4i}{L}\sum_{k>0}\sin(k(n-m))\Im\left(u_k v_k^*\right),\\(M_{BB})_{m,n}=\bra{\psi}\hat{B}_m\hat{B}_n\ket{\psi} = -\delta_{m,n} + \frac{4i}{L}\sum_{k>0}\sin(k(n-m))\Im\left(u_k v_k^*\right),\\(M_{AB})_{m,n}=\bra{\psi}\hat{A}_m\hat{B}_n\ket{\psi} = \frac{2}{L}\sum_{k>0}\cos(k(n-m))\left(|u_k|^2 - |v_k|^2\right) + \frac{4}{L}\sum_{k>0}\sin(k(n-m))\Re\left(u_k v_k^*\right),\\
    (M_{BA})_{m,n}= \bra{\psi}\hat{B}_m\hat{A}_n\ket{\psi} = - \bra{\psi}\hat{A}_n\hat{B}_m\ket{\psi} = -(M_{AB})_{n,m}
\end{gather}

All spin-spin correlators of a generic Gaussian state, including those without translational invariance, can be evaluated from Majorana correlators. The simplest one is the transverse $C^{z,z}_{m,n}$ [introduced below Eq.~\eqref{qfi_specialized}], and it reads
\begin{equation}
    C_{m,n}^{z,z} = (M_{AB})_{m,n} (M_{BA})_{m,n} - (M_{AA})_{m,n} (M_{BB})_{m,n}.
\end{equation}
This result is obtained with a straightforward application of Wick's theorem, after recalling the identity $\hat{\sigma}_j^z = \hat{\mathds{1}}-2\hat{n}_j$.
Other spin-spin correlators are instead written in terms of Pfaffians~\cite{caianiello1952,barouch1971}, namely,
\begin{gather}
    C_{m,n}^{x,x} = (-1)^{\frac{(n-m-1)(n-m)}{2}} \mathrm{Pf}\begin{pmatrix}
        (M_{BB}+\mathds{1})^{[(m,n-1),(m,n-1)]} & M_{BA}^{[(m,n-1),(m+1,n)]}\\
        M_{AB}^{[(m+1,n),(m,n-1)]} & (M_{AA}-\mathds{1})^{[(m+1,n),(m+1,n)]}
    \end{pmatrix},\\
    C_{m,n}^{y,y} = (-1)^{\frac{(n-m+1)(n-m)}{2}} \mathrm{Pf}\begin{pmatrix}
        (M_{AA}-\mathds{1})^{[(m,n-1),(m,n-1)]} & M_{AB}^{[(m,n-1),(m+1,n)]}\\
        M_{BA}^{[(m+1,n),(m,n-1)]} & (M_{BB}+\mathds{1})^{[(m+1,n),(m+1,n)]}
    \end{pmatrix},\\
    C_{m,n}^{x,y} = i(-1)^{\frac{(n-m+1)(n-m)}{2}} \mathrm{Pf}\begin{pmatrix}
        (M_{BB}+\mathds{1})^{[(m,n),(m,n)]} & M_{BA}^{[(m,n),(m+1,n-1)]}\\
        M_{AB}^{[(m+1,n-1),(m,n)]} & (M_{AA}-\mathds{1})^{[(m+1,n-1),(m+1,n-1)]}
        \end{pmatrix},
\end{gather}
where $M^{[(r_1,r_2),(c_1,c_2)]}$ indicates the submatrix of $M$ with rows from $r_1$ to $r_2$ and columns from $c_1$ to $c_2$. Notice that $\bra{\psi}\hat{\sigma}_j^x\ket{\psi}$ and $\bra{\psi}\hat{\sigma}_j^y\ket{\psi}$, appearing in the calculation of $C_{m,n}^{x,x}$ and $C_{m,n}^{y,y}$, vanish, because they are expectation values of strings of an odd number of fermionic operators that change the parity of $\hat{N}$. For the same reason, also the remaining connected correlators $C^{x,z}_{m,n}$ and $C^{y,z}_{m,n}$ are zero.

\begin{wrapfigure}{r}{0.5\textwidth}
  \begin{center}
    \includegraphics[width=0.5\textwidth]{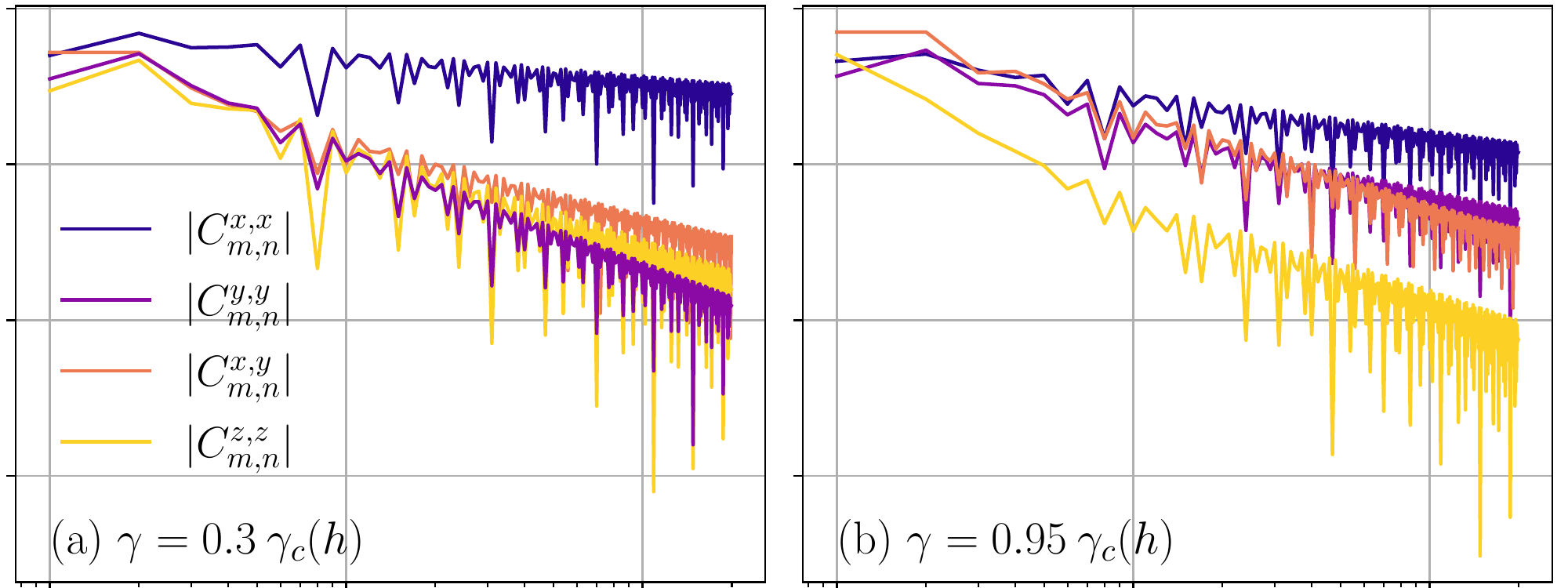}
  \end{center}
  \caption{Spin-spin correlation functions of the no-click stationary state in the logarithmic phase, for a chain of $L=8192$ spins, using $h=0.2$ and (a) $\gamma=0.3\: \gamma_c(h)$, (b) $\gamma = 0.95\: \gamma_c(h)$. The correlator with the slowest decay is $C^{xx}_j$ in both cases, but near the boundary of the phase it decreases faster than in the bulk.}
  \label{f:all_correlators}
\end{wrapfigure}

The previous formulae allow us to compute the spin-spin correlators of the vacuum state $\ket{vac}$, i.e., the stationary state of the no-click limit. As mentioned in the main text, Ref.~\cite{turkeshi2021} showed that correlation functions decay exponentially in the gapped phase, and as power laws in the gapless phase. In particular, Fig.~\ref{f:all_correlators} shows the non-vanishing correlators below the critical measurement rate. The plots are representative of what we find also for different values of $h$. Of particular interest is the longitudinal correlation function $C^{x,x}_{m,n}$, which is the slowest decaying as a function of the distance for all values of $\gamma<\gamma_c(h)$.
This explains why the QFI is maximized by a configuration of unit vectors $\mathbf{n}_j$ along the $x$ direction. We notice that the asymptotic behavior of $C^{xx}_{m,n}$ is fitted very well by the ansatz
\begin{equation}\label{c_xx_ansatz}
    C_{m,n}^{x,x}\sim \frac{\cos((\pi-k^*)|n-m|)}{|n-m|^\lambda},
\end{equation}
where the exponent $\lambda$ depends on $\gamma$. In particular, $\lambda$ appears to assume the maximum value of $\lambda\approx 0.5$ for $\gamma\ll\gamma_c(h)$, whereas it increases as $\gamma$ approaches $\gamma_c(h)$. Indeed, as argued in the main text, the exponent of the power law decay is closely related to the scaling exponent $p$ of the QFI density, and consistently Fig.~\ref{f:phase_diagram} shows a decrease of $p$ when the critical line is approached.

\begin{figure}[ht]
    \begin{subfigure}[b]{.48\textwidth}
    \centering
    \includegraphics[width=\textwidth]{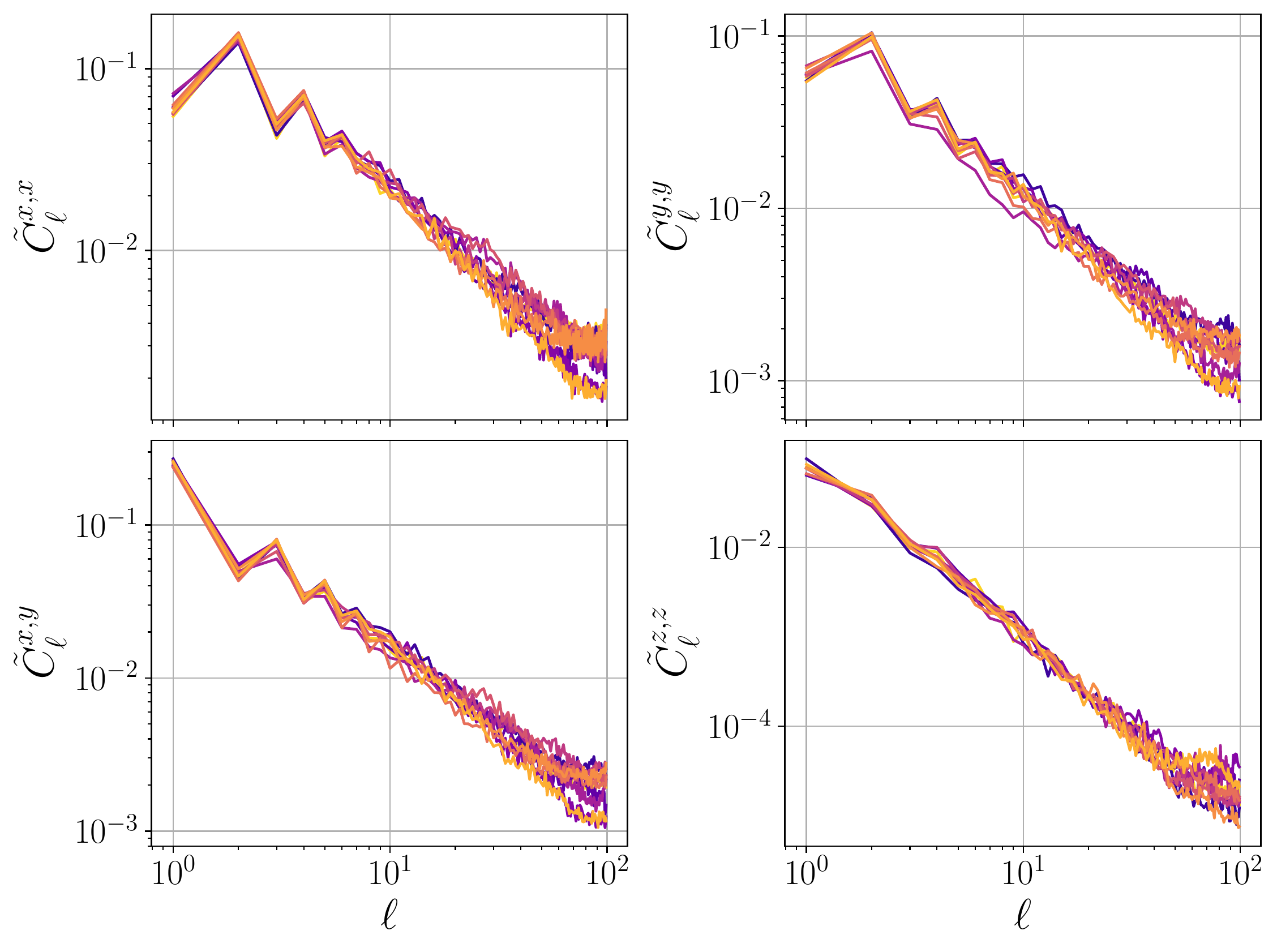}
    \caption{$\gamma=2$ (log-log scale)}
    \end{subfigure}
    \hfill
    \begin{subfigure}[b]{.48\textwidth}
    \centering
    \includegraphics[width=\textwidth]{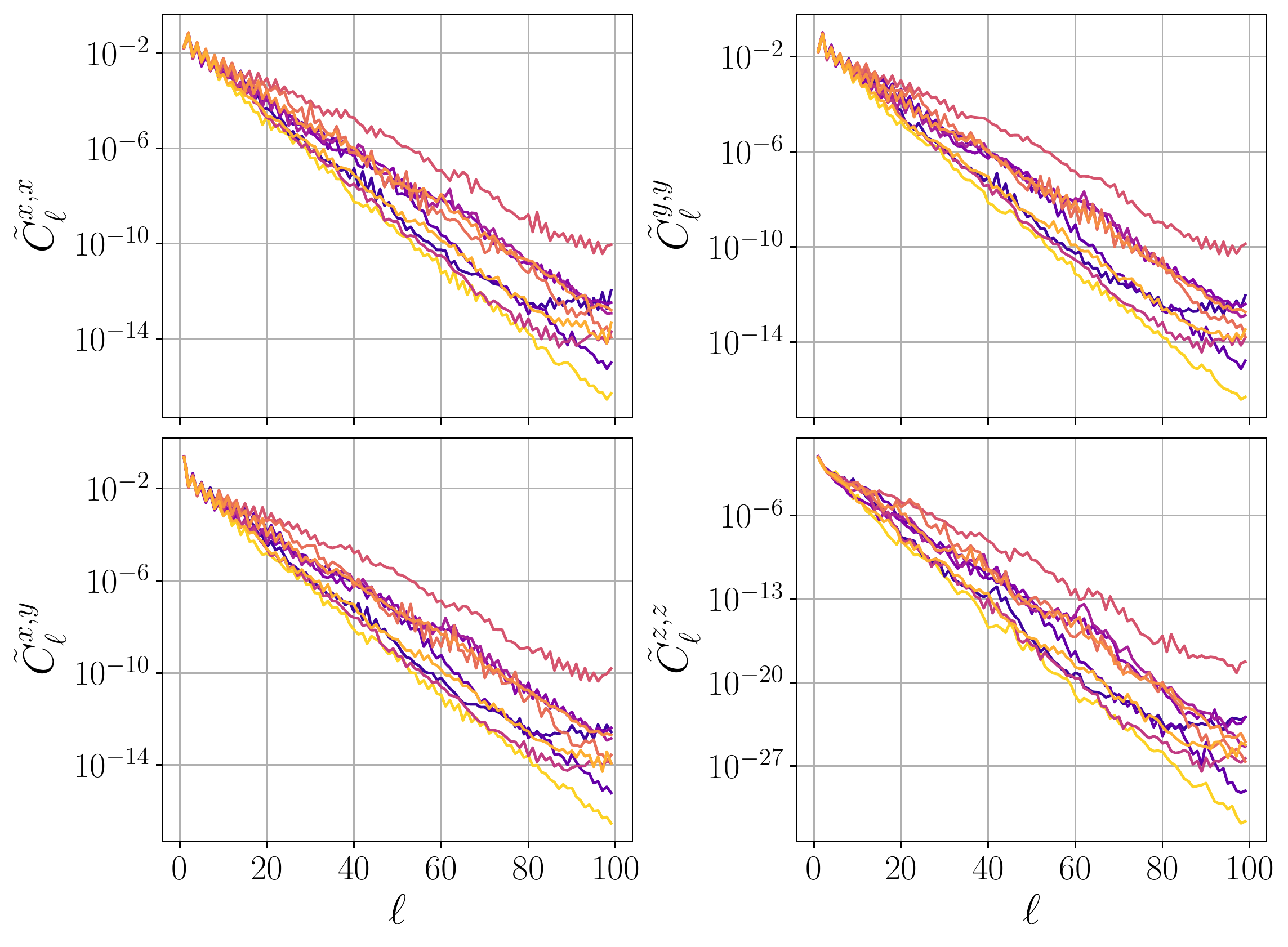}
    \caption{$\gamma=6$ (log-linear scale)}
    \end{subfigure}
    \caption{Correlators $\Tilde{C}^{\alpha,\beta}_\ell$ at long times (a) $t=250$ and (b) $t=100$, for (a) $\gamma=2$ and (b) $\gamma=6$. The plots show 10 different random realizations. The overall trend is power law for $\gamma=2$, whereas it is exponential for $\gamma=6$.}
    \label{f:long_time_correlators}
\end{figure}

As argued in the main text, the correlation functions in the long-time dynamics with jumps also feature a crossover from exponential to power laws, as in the no-click limit. Since jumps break the translational invariance of the state, we can study distance-dependent correlation functions by defining $\Tilde{C}^{\alpha,\beta}_\ell$ as in Eq.~\eqref{c_tilde}.
Figure~\ref{f:long_time_correlators} shows an example obtained for a single random realizations of quantum jumps. Establishing the exact value of $\gamma$ at which the decay shifts from exponential to power law is challenging, but our numerics indicate that it is consistent with the critical $\gamma_c$ featured by both the entanglement entropy and the QFI density. This suggests that the change in behavior of spin-spin correlators is yet another signature of the phase transition.

\begin{figure}[ht]
    \begin{subfigure}[b]{.48\textwidth}
    \centering
    \includegraphics[width=\textwidth]{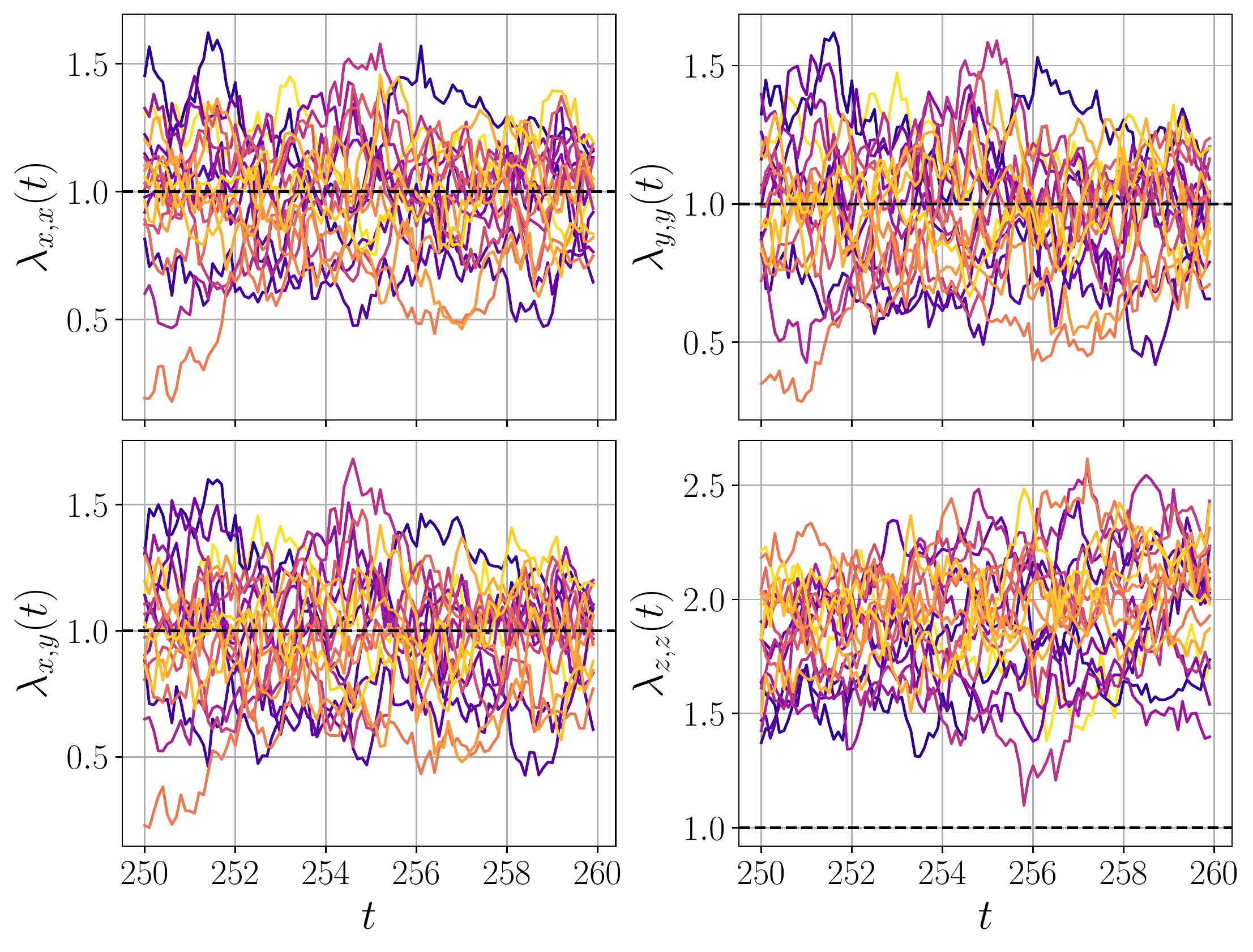}
    \caption{$\gamma=2$}
    \end{subfigure}
    \hfill
    \begin{subfigure}[b]{.48\textwidth}
    \centering
    \includegraphics[width=\textwidth]{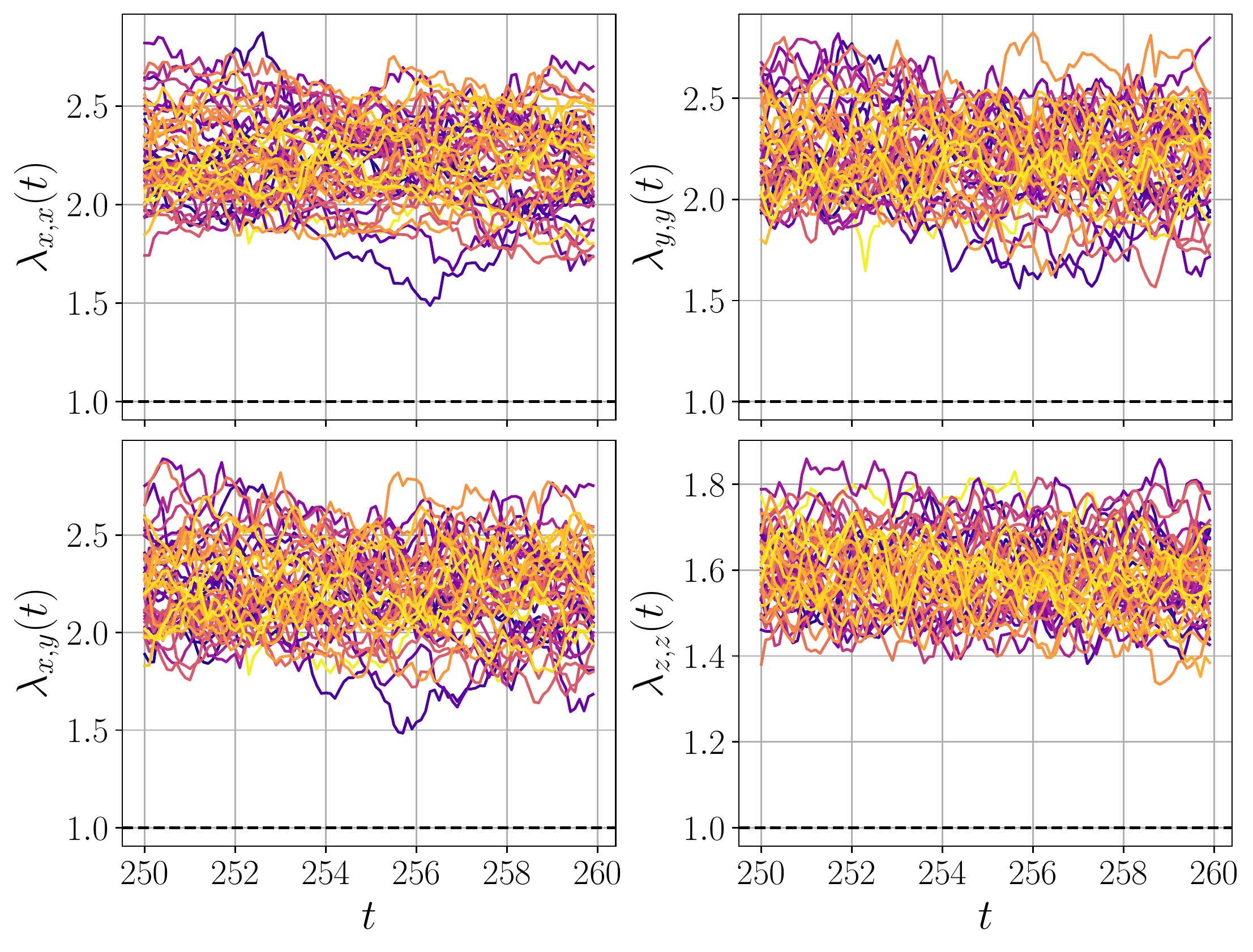}
    \caption{$\gamma=0.3$}
    \end{subfigure}
    \caption{Exponents $\lambda_{\alpha,\beta}$ of $M$ different quantum trajectories at long times $t\geq 250$, using $L=160$, $h=0.2$, (a) $\gamma=2$, $M=20$, and (b) $\gamma=0.3$, $M=50$. The exponents are extrapolated by fitting the power laws for $L=10\olddiv 60$. Many typical trajectories have exponents below $1$ when $\gamma$ is large enough, whereas no one is found for small $\gamma$}
    \label{f:traj}
\end{figure}

The values of the exponents of the power law correlators are connected to whether or not the maximal QFI density diverges in the limit of $L\to \infty$. As discussed in the main text, we study the exponents $\lambda_{\alpha,\beta}$ of $\Tilde{C}^{\alpha,\beta}_\ell$ as functions of time, because they never relax to stationary values. We present our results for $M=20$ trajectories in Fig~\ref{f:traj}a, for a value of $\gamma$ at which $\overline{f^\text{max}_Q(\infty)}$ manifests unbounded growth with $L$. Apart from $\lambda_{z,z}$, which appears to be always larger than $1$, in general we observe that other exponents fluctuate above and below $1$ at different times. Nevertheless, at any fixed time $t$, we always observe some trajectories with $\lambda_{\alpha,\beta}<1$. It is reasonable to assume that these trajectories, which are extracted randomly in our numerical implementation, are typical, and thus they are representative of a finite fraction of the whole ensemble of stochastic realizations. We believe this indicates that the average maximal QFI density indeed grows indefinitely with the system size at long times. It is worth pointing out that this conclusion relies on the additional implicit assumption that the exponents $\lambda_{\alpha,\beta}$ that we estimate for finite $L$ are representative of the true exponents found in the thermodynamic limit. Finally, for smaller values of $\gamma$, where the stationary average QFI density is intensive, we observe larger values of the exponents $\lambda_{\alpha,\beta}$, as shown in Fig.~\ref{f:traj}b. In particular, for the number $M=50$ of trajectories we explored, we never observe $\lambda_{\alpha,\beta}<1$ if $\gamma$ is small enough. This might imply that no typical trajectories that yield $f_Q^\text{max}\sim L^p$ with $p>0$ exist below a certain critical rate $\gamma_c'$. Alternatively, the fraction of such trajectories may be finite but small, making them rare to observe. With our numerical results, we are unable to establish which of the two possibilities is the correct one.

\section{Additional results}
\begin{figure}[ht]
    \begin{subfigure}[b]{.47\textwidth}
    \centering
    \includegraphics[width=\textwidth]{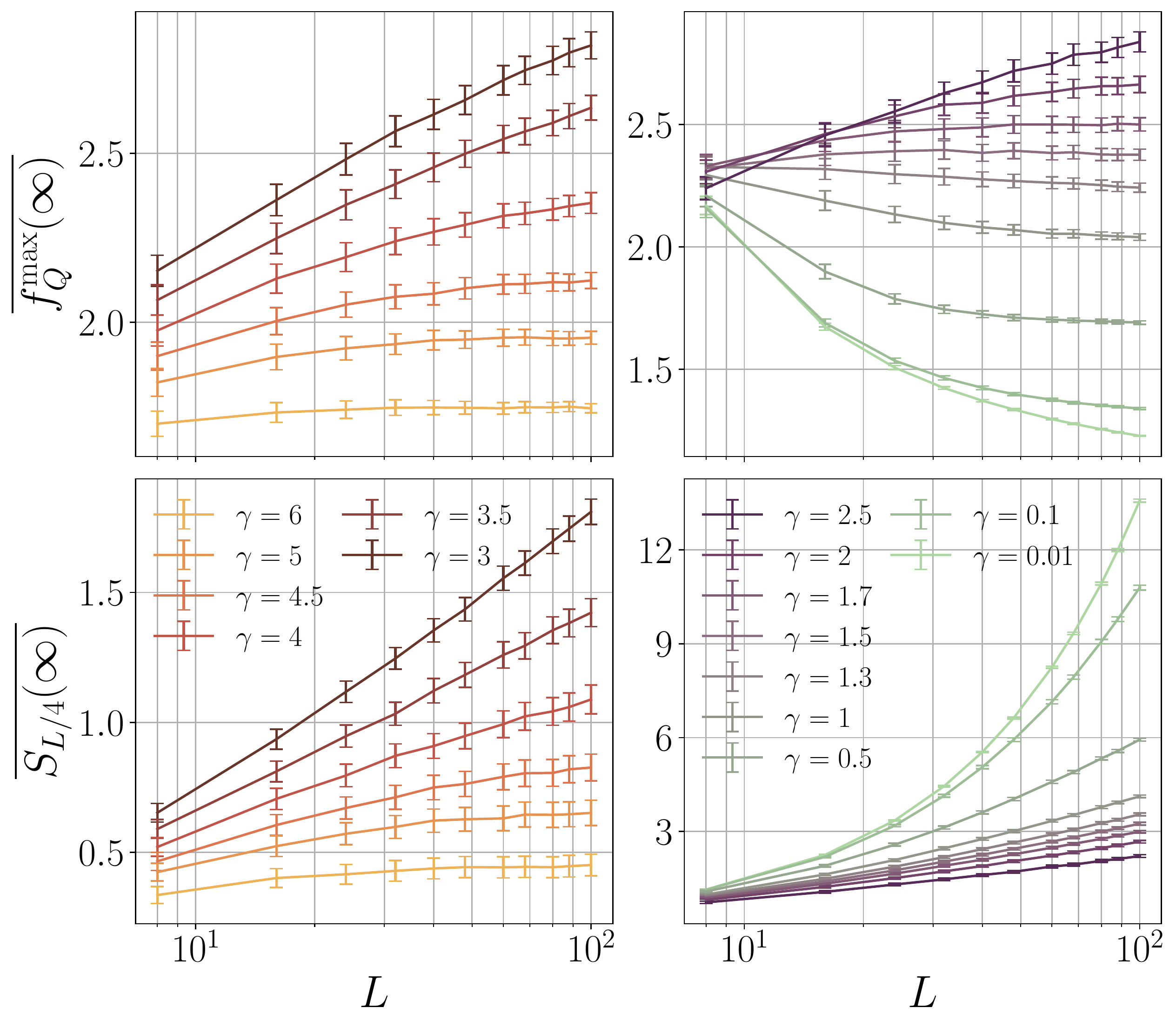}
    \caption{$h=0.7$}
    \end{subfigure}
    \hfill
    \begin{subfigure}[b]{.47\textwidth}
    \centering
    \includegraphics[width=\textwidth]{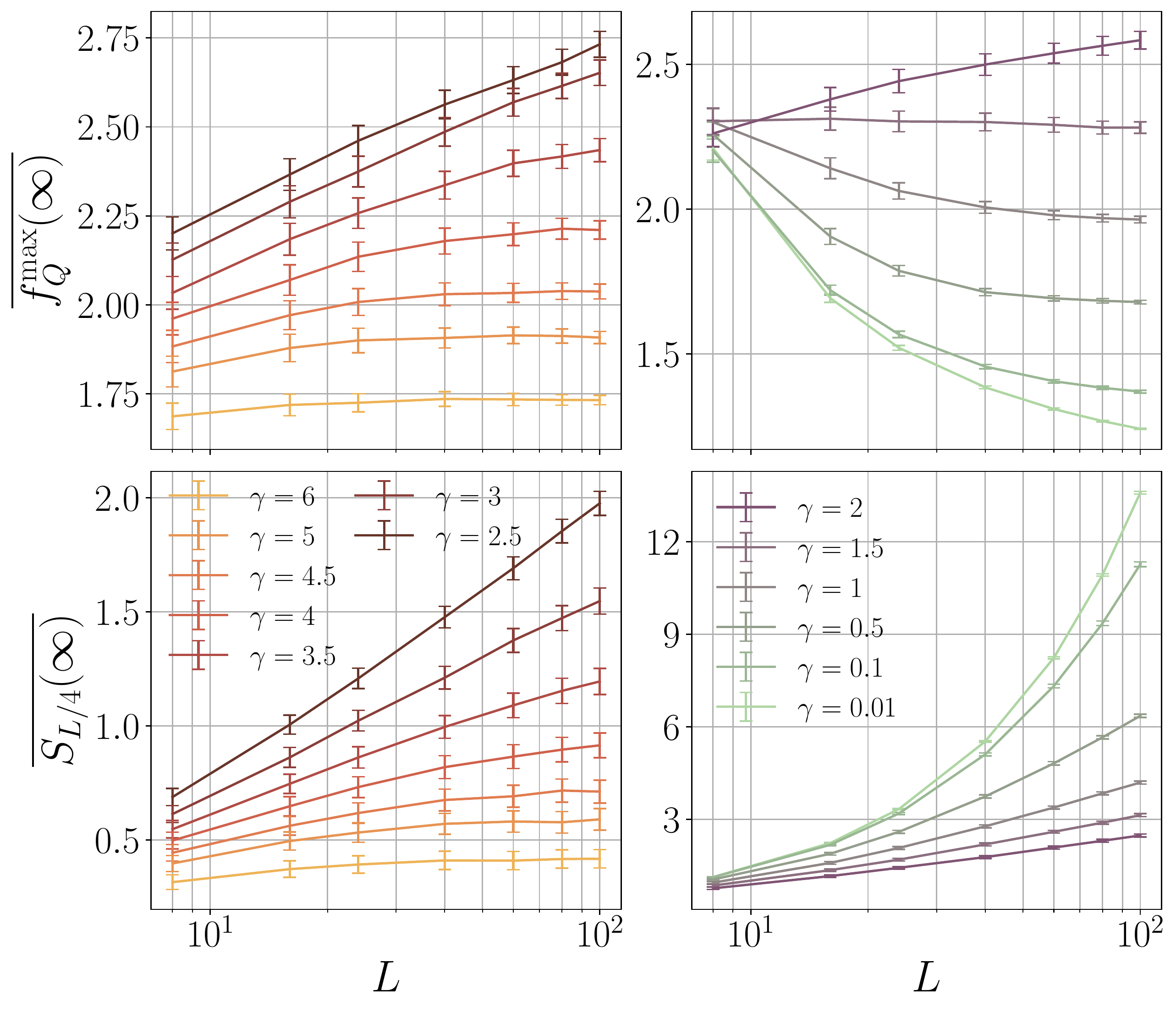}
    \caption{$h=1.2$}
    \end{subfigure}
    \caption{Disorder-averaged stationary QFI density and entanglement entropy, as in Fig.~\ref{f:qfi_scaling}, for (a) $h=0.7$ and (b) $h=1.2$.}
    \label{f:qfi_scaling_extra}
\end{figure}

In the main text we present some results on the scaling of the disorder-averaged steady-state QFI and entanglement entropy for a specific transverse field $h$. Here, in order to highlight the generality of our findings, we show numerical data for different values of $h$. Figure~\ref{f:qfi_scaling_extra} presents the same study of Fig.~\ref{f:qfi_scaling} for $h=0.7$ and $h=1.2$, respectively. Qualitatively, we observe the same behavior found for $h=0.2$. In particular, as anticipated in the main text, we find logarithmic behavior of the entanglement entropy also for $h>1$, in contrast to the numerical results presented in Ref.~\cite{turkeshi2022bis}. For both $h=0.7$ and $h=1.2$, the QFI still shows a region of unbounded growth for intermediate values of $\gamma$, whereas the entanglement entropy shifts from area to logarithm to volume law as $\gamma$ is decreased. A notable difference is that the boundaries of the three regions appear to depend on $h$. For instance, starting from the unbounded growth region and gradually decreasing $\gamma$, the QFI density seems to become intensive sooner for $h=0.7$ and $h=1.2$ than for $h=0.2$. This might indicate that the phase with divergent degree of multipartiteness shrinks in size as the transverse field gets larger.

The behavior of the long-time correlators $\Tilde{C}^{\alpha,\beta}_\ell$ also remains analogous for the other values of $h$. In detail, the correlation functions still shift from exponential to power laws as $\gamma$ is reduced, and the unbounded growth of the QFI density can still be attributed to slowly-decaying correlators with $\lambda_{\alpha,\beta}<1$.

\end{document}